\documentclass[twocolumn]{revtex4}

\newcommand{\hg}{ }

\usepackage{amssymb}
\usepackage{graphicx}
\usepackage{epstopdf}
\usepackage{colordvi}
\usepackage{url}

\begin{document}

\title{Negative Differential Resistance due to Nonlinearities in Single and Stacked Josephson Junctions}
\author{
G. Filatrella$^1$,  V. Pierro$^2$,  N.F. Pedersen$^3$,  and  M.P. S{\o}rensen$^3$\\
$^1$ Department of Science and Technology, University of Sannio, I-82100, Benevento, Italy\\
$^2$ Department of Engineering, University of Sannio, I-82100, Benevento, Italy \\
$^3$ Department of Applied Mathematics and Computer Science, Technical University of Denmark, 2800 Kgs. Lyngby, Denmark
}
\date{\today}

\begin{abstract}
Josephson junction systems with a negative differential resistance play an essential role for applications.
As a well known example, long Josephson junctions of the BSCCO type have been considered as a source of THz radiation in recent experiments. 
Numerical results for the dynamics of the fluxon system have demonstrated that a cavity induced negative differential resistance plays a crucial role for the emission of electromagnetic radiation. 
We  consider the case of a negative differential resistance region in the McCumber curve itself of a single junction and found that it has an effect on the emission of electromagnetic radiation. 
Two different shapes of negative differential resistance region are considered and we found it is essential to distinguish between current bias and voltage bias. 
\end{abstract}
\maketitle

\section{Introduction}
\label{introductionj}
Fluxon  dynamics in long Josephson junctions is a topic of strong interest due to its rich nonlinear properties and applications in fast electronics, in particular as a radiation source at high
frequencies \cite{Ozyuzer07,Wang10,Koshelets00,Kashiwagi12}. 
An extension of that system is to form a metamaterial by stacking several Josephson  junctions on top of each other, which are modeled by $N$ coupled partial differential
equations.  

{   Such superconductors are employed  in a variety of devices (stacks, flux-flow devices, and synchronized arrays) and are capable of power emission in the range 0.5-1 THz. Integration in arrays could give an improvement in the power performances, above $100 \mu W$ \cite{Welp13}. Practical applications are especially in the field of bio-sensing, nondestructive testing and high speed wireless communications \cite{Welp13}. For such reasons we aim to understand if some simple mechanism is at work in all these devices.  }
Such a system is used as a model for high temperature superconductors
of the BSCCO type \cite{Tachiki09,Welp13}. In this communication we go one step further in complexity and include results on a nonlinear behavior of the shunt resistor, giving rise to features similar to stacked Josephson junctions coupled to a cavity \cite{Madsen08,Madsen10,Tachiki11,Tsujimoto10}.
Such a model is needed in order to understand and interpret the experimental measurements. 
For frequencies in the GHz or even THz range, either an intrinsic or an external cavity is needed to enhance the radiated power to useful levels. 
Figure \ref{Fig1}a shows qualitatively the appearance of a nonlinear current-voltage (IV) curve for the quasi particle tunneling in the Josephson junction. 
The particular form of the IV curve resembles a distorted N and
hence we refer to this particular form as $N$-shaped IV curve. 
Note that the quasi particle
tunnel current is a unique function of the applied voltage, but the inverse function is not unique. 
Similarly, Fig. \ref{Fig1}b depicts an IV curve, which is shaped as a distorted $S$ ($S$-shaped IV curve). 
In this latter case the voltage is a unique function of the current. In general the nonlinear
behavior leading to a negative differential resistance (NDR) of Josephson junctions plays a key role for applications. 
An example is a parametric amplifier or a radiation source at high frequencies \cite{Soerensen79,Wahlsten77,Pedersen73}. 
Examples of NDR are numerous: (i) Josephson junction with a cavity \cite{Barbara99,Filatrella03},
(ii) backbending of the BSCCO energy gap \cite{Welp13}, (iii) structure at the energy gap difference if the junction consists of two different superconductors, (iv) in connection with Fiske steps, zero field steps \cite{Cirillo98}  and even in rf-induced steps \cite{Pedersen80}. 
In some cases of NDR a nonlinear region that cannot be explained shows up in the IV curve \cite{Kadowaki}.
The two qualitatively different cases of a nonlinear differential resistance, referred to as
$N$-shaped and $S$-shaped  regions, observed in experiments will be
discussed in the next sections. 
We mention that besides the NDR at finite voltages, also absolute negative resistance \cite{Nagel08} and negative input resistance at zero voltages \cite{Pedersen76} have been reported. 

{     In this work we want to emphasize the role of nonlinearities and to show that even a very simple model can give rise to an interesting power emission profile. However, our model, being a high frequency model, cannot capture effects at low timescale, such as thermal effects \cite{Wang10,Yurgens11,Li12,Gross13}. }
We discuss below various examples of a negative differential resistance in Josephson junctions.

\section{Negative differential resistance at finite voltages}
\label{ndrfinite}
Josephson junctions come in a variety of forms with different properties but the same generic behavior.
Some examples are: (i) The traditional low temperature ($T_c$) Josephson junction with qualitatively different behaviors depending on the dimensions of the junction, (ii) high $T_c$ intrinsic Josephson junctions that are typically described as a stack of long Josephson junctions leading to coupled sine-Gordon equations and (iii) point contacts and microbridges that are the easiest to describe mathematically. 
Some features are generic, like the Josephson voltage to frequency relation, the supercurrent, the energy gap etc.
In all cases we may have a coupling to a cavity either intrinsic (internal) or external, which of course complicates the mathematics but may be important for applications.

\begin{figure}[tbp]
\begin{center}
\includegraphics[width=7cm]{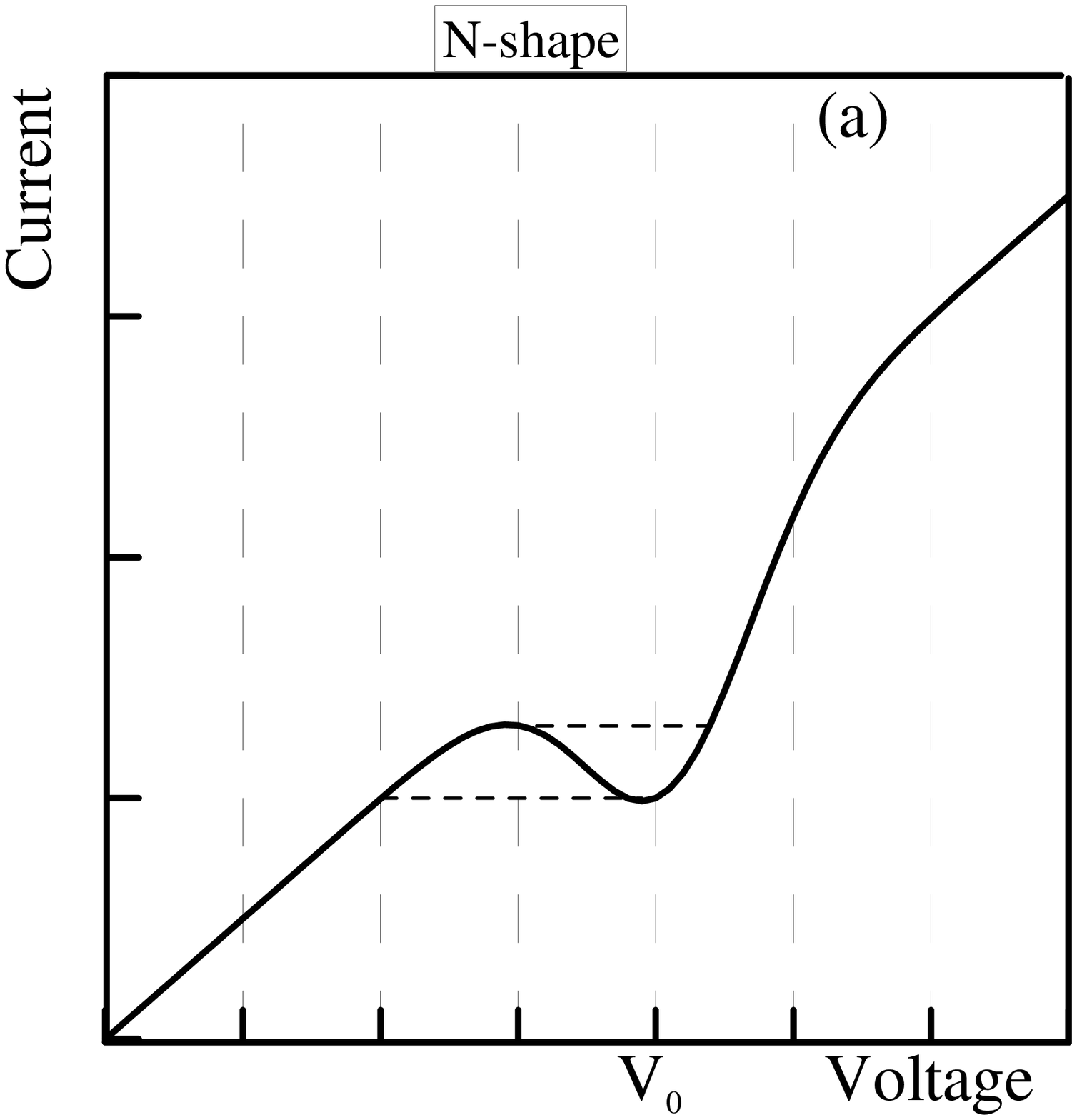}\\
 \vspace{5mm}
\includegraphics[width=7cm]{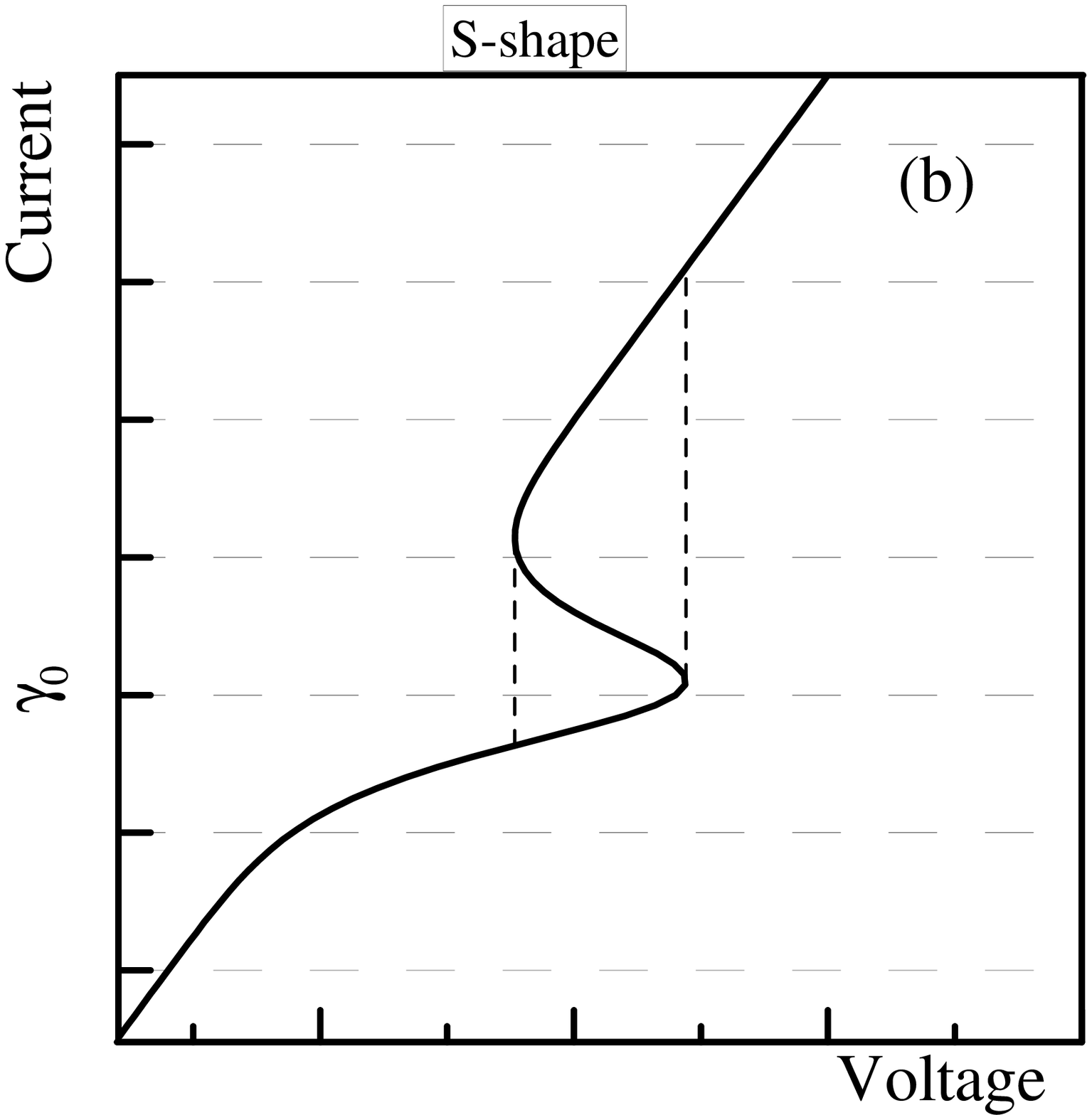}
\caption{Negative differential resistance for Josephson junctions and some semiconductors.
a) $N$-shape, and b) $S$-shape. The lines are the IV curves (showing hysteresis). The dashed lines indicate the bias, voltage for the $N$-shape resistor, current for the $S$-shape case. The short-dashed lines show the switching from a branch to another when the bias is reversed /current bias for the $N$-shape, voltage bias for the $S$-shape. Finally, $V_0$ and $\gamma_0$ represent the dip of the negative differential
resistor, see Sections \ref{sshape} and \ref{figure}.}
\label{Fig1}
\end{center}
\end{figure}

The two different cases of  negative differential resistance – $N$-shaped and $S$-shaped – discussed here are shown in Fig. \ref{Fig1}.

\begin{figure}[tbp]
\begin{center}
\includegraphics[height=6cm,angle=0]{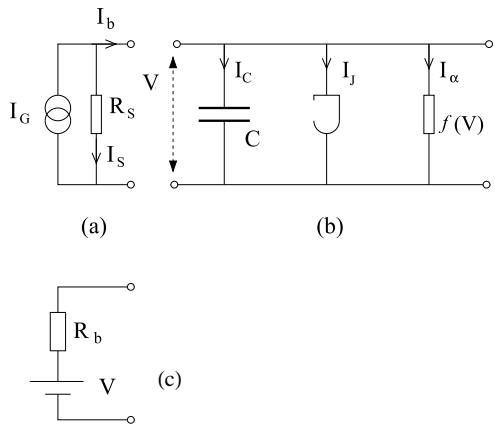}
\caption{The Josephson junction model with a current bias circuit (a) or a voltage bias circuit (c).
The center section (b) is a resistively shunted Josephson circuit with a nonlinear resistor $I_{\alpha} = f(V)$.
}
\label{Fig2}
\end{center}
\end{figure}

Fig. \ref{Fig1}a shows schematically both the NDR IV curve of a semiconductor Gunn diode \cite{wikipedia1} (which is used as a microwave source) as well as that of a Josephson junction coupled to a cavity \cite{Madsen10}. This type of IV curve is sometimes referred to as an $N$-shaped IV curve \cite{wikipedia1}. 

{ The analogy with Gunn diode is purely hypothetical. We speculate that the NDR IV characteristic, even if of very different nature, could be similar to Josephson junction NDR at a phenomenological level. This is only an observation, and we do not claim that the constitutive physics is the same.} 
In order to measure the NDR part of the IV curve in 
Fig. \ref{Fig1}a, we must have a constant voltage bias source. With a constant current bias source we will see hysteresis when increasing and decreasing bias current (dashed line) . For the Josephson junction the cavity may typically look as the hysteretic curve if the junction is current biased or as the continuous curve (full line) when voltage bias. In most cases the experimental curves for Josephson junctions are current biased although voltage bias can also occur \cite{Gokhfeld07}.

Figure \ref{Fig1}b shows another type of NDR curve referred to as $S$-shaped. Here a constant current bias source is needed to trace out the full curve.
In reference \cite{Ozyuzer07} a NDR region
qualitatively similar to that in Fig. \ref{Fig1}b is observed together with radiation emission.
Since additionally in \cite{Ozyuzer07} radiation emission is also observed in a part
of the IV curve that looks qualitatively similar to that in Fig. \ref{Fig1}a, it
is tempting to link the radiation emission with the NDR. We note that a Josephson
junction is nonlinear up to very high frequencies (THz), implying that it's negative resistance may have important high frequency applications: examples are (i) a high frequency parametric amplifier and (ii) a generator for high frequency electromagnetic radiation. 
We note from Fig. \ref{Fig1} that we have 4 different combinations of the
bias and $N/S$ nonlinearity. In the general case the bias is neither current nor voltage biased but has a load line that is neither vertical nor horizontal.
However, a pure voltage bias is not conceivable for a superconducting element, and therefore there is not a full symmetry between the two cases, as can be seen in the following.

For intrinsic Josephson junctions the NDR qualitatively shown in Fig. \ref{Fig1}b comes from the
well known back bending of the IV curve near the energy gap due to heating.
There are other shapes than $S$-shaped and $N$-shaped, an example from semiconductors is the
$Z$-shape 

A diagram showing the model we are discussing is depicted in Fig. \ref{Fig2}.
Here the left part of the figure shows the two
different bias circuits, the middle part (b) shows the Josephson junction model with the nonlinear resistor and the capacitor. 

{  We do not hypothesize on the physical origin responsible for the NDR. Instead, a phenomenological IV characteristic of the effective shunt encompasses the behavior of the interaction with an external load. Shortly, we adopt the following logic: the observed IV of loaded stacks exhibits a NDR, therefore we introduce a nonlinear effective shunt resistor. }

For the bias circuit we distinguish between (a) current bias and (c) voltage bias. 
In the case of current bias the shunt resistance $R_S$ is large and for the voltage bias case the bias resistor $R_b$ is small. 
We underline that $R_b=0$ leads to an unphysical result for a superconducting
element as the Josephson junction. In fact a dc voltage produces an ac current
$I_0 \sin(\omega t + \varphi_0)$ through the Josephson junction element, while the capa\-ci\-tor remains inactive and the resistor current is constant.

{  The actual bias at high frequency is more complicated than elements (a) and (c) of Fig. \ref{Fig2} \cite{Likharev86}, and we use such very simple schemes for sake of simplicity. We notice that in experiments a partial voltage bias has been employed (e.g., \cite{Ozyuzer07,Welp13}), and a  voltage bias is sometimes used in modeling (e.g.,\cite{Gokhfeld07}); the interface between the voltage source and the mesa  at THz frequencies leads to a very difficult full wave electromagnetic problem. }

In the following we shall set up a mathematical model for the Josephson junction circuits depicted
in Fig. \ref{Fig2}. The voltage across the Josephson junction is denoted $V$=$V(t)$ and it depends on time $t$.
The Cooper pair phase difference across the Josephson barrier is $\phi$=$\phi(t)$ and the Josephson
voltage relation reads $V$=$[\hbar/(2e)] d\phi/dt$. The electron charge is $e$ and $\hbar$ denotes Planck's constant divided by
$2\pi$. .
We model the $N$-type NDR characterized by a nonlinear
conductance $G(V)$ shaped as a Gaussian function

\begin{equation}
\label{G}
G(V) =
G_0 \left\{ 1 + \alpha_1 \exp \left[-\frac{\left(V-V_0\right) ^2}{ \delta} \right] \right\} \; .
\end{equation}

\noindent Here $G_0$ is the linear quasi particle conductance (related to the quasi particle resistance $R_j = 1/G_0$) and $\alpha_1$ describes the nonlinear part of the conductance.

The parameter $V_0$ position the Gaussian function along the $V$-axis and $\delta$
is a measure of the Gaussian width.

{   The Gaussian model is phenomenological -- it is only a way to create the negative differential resistance. In fact, we have also tried with other functional shapes (e.g. polynomial), obtaining the same qualitative result: an increase of the power associated with the NDR branch, as in \cite{Ozyuzer07,Welp13}. A detailed comparison that could give a quantitative evaluation of the parameters $\alpha_1$ and $\delta$ is out of the scope of this paper, which aims to a phenomenological description. }

 The quasi particle tunnel current $I_R$ is then given by
\begin{equation}\label{Iqp}
  I_R(V)= G(V) V \; .
\end{equation}

\noindent In Fig. \ref{Fig2} $I_b$ denotes the bias current supplied by the current source
in the circuit (a) or voltage source in the circuit (d), $I_C$ is the displacement current
through the capacitor and the Josephson tunnel current is denoted $I_J$ = $I_0 \sin(\phi)$,
where $I_0$ is the Josephson critical current.
The current balance becomes $I_C+I_R+I_J=I_b$. 
Inserting the Josephson voltage relation and Eq. (\ref{Iqp}) into the current balance equation gives

\begin{equation}\label{RSJ1}
C \frac{\hbar}{2e} \frac{d^2 \phi}{dt^2} + \frac{\hbar}{2e} G \left( \frac{d \phi}{dt} \right) \frac{d \phi}{dt}
+ I_0 \sin(\phi) = I_b \; .
\end{equation}

\noindent  The above Eq. (\ref{RSJ1}) can be transformed into normalized units using the scaling
$t=k_t \tau$, with $k_t=\sqrt{\hbar C}/\sqrt{2 e I_0}$, thereby we obtain

\begin{equation}
\label{RSJnorm}
\frac{d^2 \phi}{d \tau^2} + \alpha \left(\frac{d\phi}{d \tau}\right) \frac{d \phi}{d\tau} + \sin(\phi) = \gamma_b \; .
\end{equation}

\noindent Here the nonlinear conductance $G(V)$ has been normalized and leads to
the nonlinear dissipation term with $\alpha_0=(1/R_j)(\hbar/2eCI_0)^{1/2}$ in

\begin{equation}
\label{Gnorm}
\alpha\left(\frac{d \phi}{d \tau}\right) =
\alpha_0  \left\{ 1 + \alpha_1 \exp \left[ - \frac{1}{\Delta} \left(\frac{d \phi}{d \tau}-v_0\right) ^2 \right]  \right\} \; .
\end{equation}

\noindent In the above expression $\Delta=\delta/V_N^2$, $V_N=\alpha_0R_jI_0$, and $v_0=V_0/V_N$ is the normalized voltage. The equations reduce to the usual RSJ model for $\alpha_1=0$.

Model Eq.(\ref{RSJnorm}) is a phenomenological equation for the NDR observed in the full model \cite{Madsen10}: 

\begin{eqnarray}
\frac{\partial^2\phi}{\partial x^2}&=& {\mathbf S}{\mathbf J_z }
\label{stacka}  \\
J^i_z & =& \frac{\partial^2 \phi ^i}{\partial \tau^2}+ \alpha \frac{\partial \phi^i}{\partial \tau} + \sin \phi^i - \gamma_b,
\label{stackb}
\end{eqnarray}

\noindent where ${\mathbf S}$ is the tridiagonal matrix that describes the coupling among the superconducting layers \cite{Sakai93}. The matrix reads $1$ in the diagonal, and the coupling constant $S$ otherwise. In the following we will adopt the phenomenological model (\ref{RSJnorm}) that qualitatively describes the NDR obtained by the full model (\ref{stacka},\ref{stackb}) (see Ref.  \cite{Madsen10}).

\begin{figure}[hbp]
\begin{center}
\includegraphics[width=6cm]{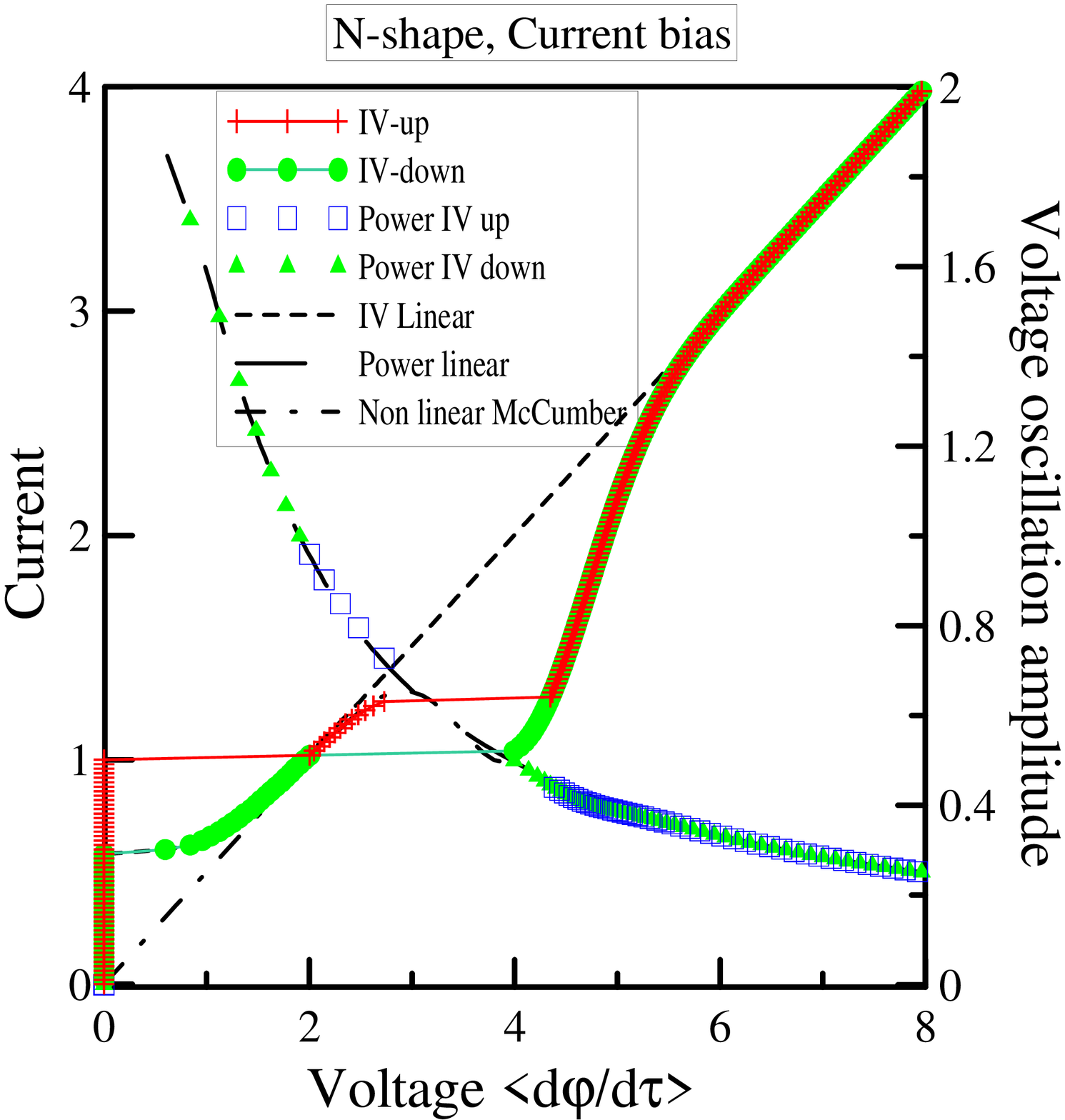}
\caption{ IV curve and voltage oscillation for a Josephson junction coupled to a $N$-type
resistor and current biased. 
Crosses (red symbols) indicate increasing bias current and circles (green symbols)
depict the case of decreasing bias current.
The available power is estimated through the maximum voltage oscillations. Parameters of the simulations are: $\alpha=0.5$, $\Delta=0.1$, $v_0=4$, $\alpha_1=-0.5$ . }
\label{FIG_Ibias_Nshape}
\vspace{5mm}
\includegraphics[width=6cm]{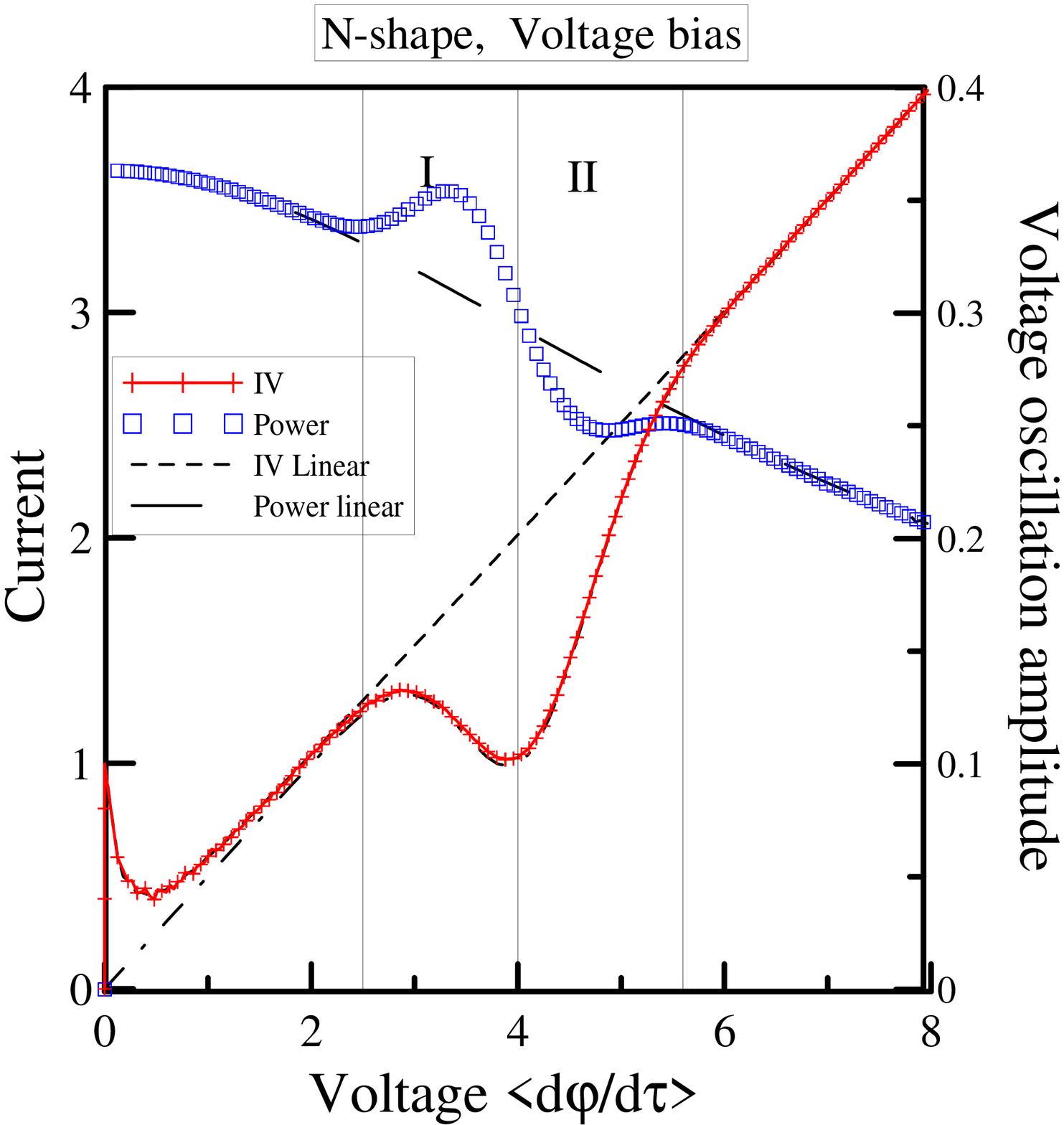}
\caption{ Non hysteretic voltage biased IV curve and voltage oscillations for a Josephson junction coupled to a $N$-type resistor. 
Crosses (red symbols) indicate increasing and decreasing bias current.
The available power is estimated through the maximum voltage oscillations. 
Parameters of the simulations are  $\alpha=0.5$, $v_0=4$, $\alpha_1=-0.5$, $\Delta=0.1$, $r_b=0.1$.}
\label{FIG_Vbias_Nshape}
\end{center}
\end{figure}

\section{$N$-shaped resistor}
\label{nshape}

Inserting a $N$-shaped resistor in $\alpha(d\phi/d\tau)$, see Eq.(\ref{Gnorm}),  leads to current voltage curves of the form depicted in Fig. \ref{FIG_Ibias_Nshape}. The model for the $N$-shaped resistor can also be used in the case of voltage bias, where the normalized current reads (see Fig. \ref{Fig2}):

\begin{equation}
\label{Gbias}
\gamma_b=\frac{\alpha_0}{r_b} \left( \frac{d\phi}{d\tau} - \frac{V}{V_N} \right),
\end{equation}

\noindent where $r_b$ is the normalized resistance $R_b$ of Fig. \ref{Fig2}c. The expression in (\ref{Gbias}) can be inserted into the system equation (\ref{RSJnorm}), which reduces to the current bias when $\alpha_0 \rightarrow 0$ and $V/V_N\rightarrow \infty$.
{ In this setup we consider open circuit boundary conditions. More realistic boundary conditions as an RLC circuit will be discussed elsewhere \cite{unpublished_C}. }

{\hg
In Fig. \ref{FIG_Ibias_Nshape} the available power is estimated through the oscillations of the Josephson voltage. 
We underline that in this paper we refer to  {\it available power}, or shortly {\it power} (also in the figure labels), as the maximum power that could be extracted supposing a perfect matching with the external device. The optimization of this important effect is relevant for the application, but in this paper we focus on the existence of a general effect due to NDR. }
  The power estimate is done recording the JJ waveforms and extracting the difference between the maximum and the minimum, that it is proportional to the mean square amplitude in the sinusoidal regime. The features of the power estimate are generic and consistent with a full Fourier analysis of the waveform.  The nonsinusoidal response of a single JJ at low bias is enhanced and results in a high harmonic content at low bias, close to the return point on the IV. 
 {\hg The voltage oscillations  seems to go towards infinity at low dc voltage. 
As it is well known  \cite{Likharev86}, the actual power is peaked at some (low) frequency and goes towards zero at high frequencies (due to capacitance ) and low frequencies (due to inductance). It is also known that there is a phase change in the ac-voltage (respect to the ac current) as we decrease bias current, such that $I\cdot V$ will not increase. 

Moreover, in our calculations we only find the rising branch that is stable, and that abruptly switches to zero voltage at a finite level of amplitude (corresponding to a finite current before the switch to the Josephson supercurrent). 
Applicationwise the most interesting region is above the peak when the voltage, and hence the frequency of the emitted radiation, span the THz region.
}

In summary, from Figs. \ref{FIG_Ibias_Nshape}, \ref{FIG_Vbias_Nshape} it is evident that:
(i) The nonlinear McCumber branch causes a hysteretic behavior.
(ii) The output power, i.e. the amplitude of the Josephson voltage oscillations, is nonlinear in the region of McCumber nonlinearity.
(iii) The negative branch of the nonlinearity (region I in Fig. \ref{FIG_Vbias_Nshape}, corresponds to a power higher than the power obtained from a linear McCumber resistor. This is only evident in the case of voltage bias, for the nonlinear part is unstable with a current bias.
(iv) The reverse is true for the positive branch of the nonlinearity (region II in 
Fig. \ref{FIG_Vbias_Nshape}), where the available power is lower than the power obtained from a linear McCumber resistor.

\section{$S$-shaped resistor}
\label{sshape}

An $S$-type NDR may also be found in a Josephson junction. The most obvious example is the back bending of the IV curve in an intrinsic Josephson junction at the energy gap \cite{Welp13}. This case is mathematically more complex than the $N$-type. For the $N$-type case we noticed that to have a continuous IV curve with NDR we should have voltage bias   (Fig. \ref{Fig1}a). Most experiments on Josephson junctions are current biased and give rise to hysteresis (See Fig.  \ref{FIG_Ibias_Nshape}). For $S$-type nonlinearity the situation is the inverse. The continuous IV curve is obtained with current bias whereas with voltage bias switching will result (Fig \ref{Fig1}b).

If we consider a current biased Josephson junction circuit with an $S$-type nonlinearity (see 
Fig. \ref{Fig1}b) we obtain for a nonlinear resistor:

\begin{eqnarray}
\label{GS}
G(I_R) &=&
G_0  \left\{ 1 + \alpha_1 \exp\left[-\frac{(I_R-\tilde{I})^2}{\delta} \right] \right\} \\
I_R &=&\frac{\hbar}{2e}G(I_R)\frac{d\phi}{d t}.
\label{IRS}
\end{eqnarray}
\begin{figure}[tbp]
\begin{center}
\includegraphics[width=7cm]{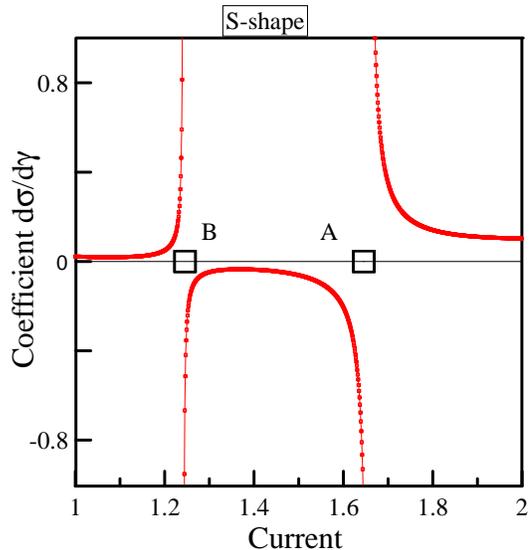}
\caption{ Sketch of the coefficient of Eq. (\ref{RSJ_S_sigma}) for an $S$-type NDR model.}
\label{Lip}
\end{center}
\end{figure}

\noindent  That reduces to the linear case for $\alpha_1 = 0$. As usual with the current balance one gets the set of equations:

\begin{eqnarray}
\label{cbalance}
I_R+I_C+I_j = I_b  \nonumber \\
I_R =\frac{\hbar}{2e}G(I_R)\frac{d\phi}{d t}.
\end{eqnarray}

\noindent Inserting the expressions for the current through the capacitor ($I_C$) and the
 Josephson element ($I_J$) one retrieves the second order set of differential
 equations:

\begin{eqnarray}
\label{RSJ_S}
I_R+C \frac{\hbar}{2e} \frac{d^2 \phi}{dt^2} + I_0 \sin(\phi)  = I_b \nonumber \\
I_R =\frac{\hbar}{2e}G(I_R)\frac{d\phi}{d t}  \; .
\end{eqnarray}

\noindent That can be cast in normalized units (with the standard definitions of Josephson angular velocity $\omega_0$ and dissipation $\alpha_0$) in a system of two equations:

\begin{eqnarray}
\label{RSJ_S_n}
\gamma_R+ \frac{d^2 \phi}{d\tau^2} +  \sin(\phi)  = \gamma_b \nonumber \\
\gamma_R = \alpha \left(\gamma_R \right)\frac{d\phi}{d\tau}  \; .
\end{eqnarray}

\noindent Here the nonlinear conductance $G(I_R)$ has been normalized and
gives rise to  nonlinear dissipation $\alpha(\gamma_R)$:

\begin{equation}
\label{Gnorm_S}
\alpha \left( \gamma_R  \right) =
\alpha_0  \left\{ 1 + \alpha_1 \exp \left[ -  \frac {\left( \gamma_R- \gamma_0 \right) ^2}{\Delta'} \right]  \right\} \; ,
\end{equation}

\noindent where $\Delta' = \gamma_0/I_0^2$. The equations can be finally cast in normal form with the variables $\phi$ and $\gamma_R$:

\begin{eqnarray}
\label{RSJ_S_normal}
 \frac{d\gamma_R}{d\tau} &=& \left\{
\frac{\alpha^2 \left(  \gamma_R  \right) }{\alpha\left(  \gamma_R  \right) - \gamma_R \frac{d\alpha\left(  \gamma_R  \right)}{d\gamma_R}}
\right \}
 \left[   \gamma_b -\gamma_R -  \sin(\phi)  \right]  \nonumber \\
 \frac{d\phi}{d\tau} &=& \frac{\gamma_R}{\alpha\left(  \gamma_R  \right)}  \; .
\end{eqnarray}

\noindent The equations reduce to the usual RSJ model for $\alpha_1=0$:

\begin{figure}[tbp]
\begin{center}
\includegraphics[width=7cm]{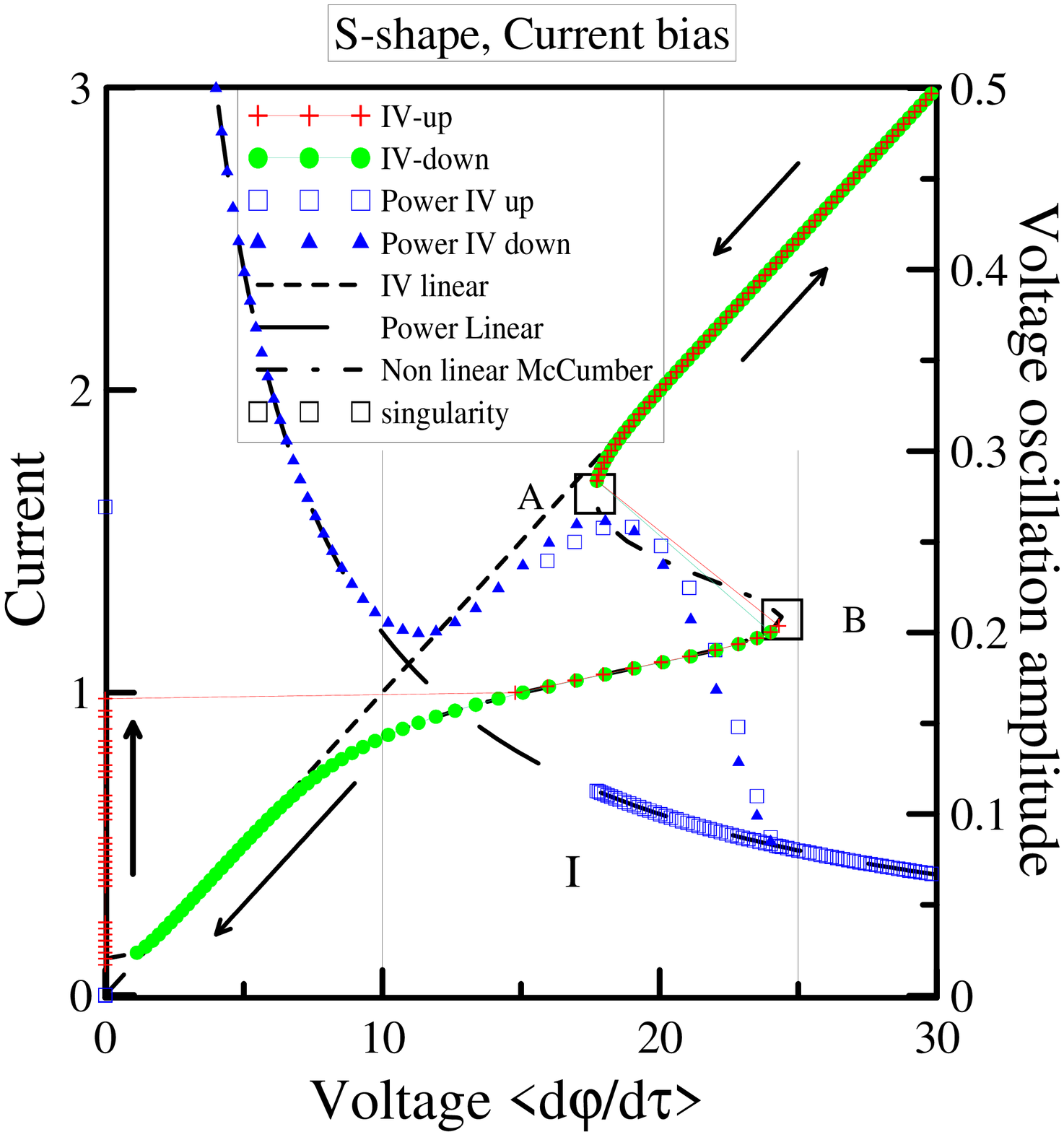}
\caption{ IV curve and voltage oscillations for a Josephson junction coupled to a $S$-type
resistor and current biased. 
Crosses (red symbols) indicate increasing
bias current and circles (green symbols)
depict the case of decreasing bias current. The available power is estimated through the maximum voltage oscillations. Parameter of the simulations are: $\alpha=0.1$, $\gamma_0=1.2$, $\Delta'=0.1$, $\alpha_1=-0.5$.}
\label{FIG_Ibias_Sshape}
\vspace{5mm}
\includegraphics[width=7cm]{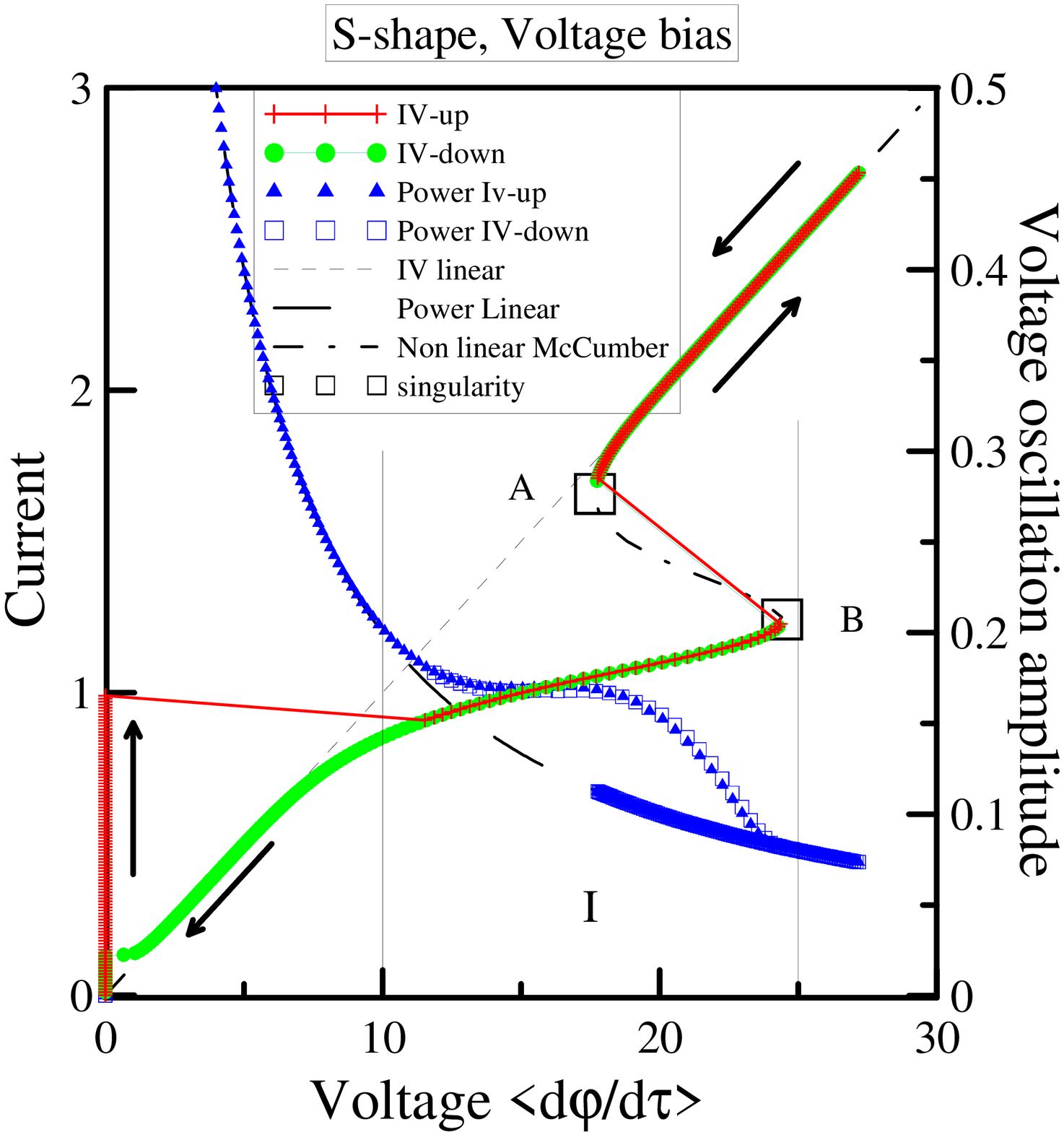}
\caption{ IV curve and voltage oscillations for a Josephson junction coupled to a $S$-type
resistor and voltage biased. 
Crosses (red symbols) indicate increasing
bias current and circles (green symbols)
depict the case of decreasing bias current. The available power is estimated through the maximum voltage oscillations. Parameter of the simulations are: $\alpha=0.1$, $\gamma_0=1.2$, $\Delta'=0.1$, $\alpha_1=-0.5$, $r_b=0.1$. }
\label{FIG_Vbias_Sshape}
\end{center}
\end{figure}

The dissipation is employed to write the system of differential equations for the variables $\phi$ and $\gamma_R$ that is true for any functional dependence of the nonlinear dissipation upon the resistor current  $\gamma_R$

\begin{eqnarray}
\label{RSJ_S_general}
 \frac{d\gamma_R}{d\tau} &=& \left\{  \frac{d}{d\gamma_R} \frac{\gamma_R}{\alpha\left(  \gamma_R  \right) } \right \}^{-1}  \left[   \gamma_b -\gamma_R -  \sin(\phi)  \right]  \nonumber \\
 \frac{d\phi}{d\tau} &=& \frac{\gamma_R}{\alpha\left(  \gamma_R  \right)  }  \; .
\end{eqnarray}

\noindent In general, for any $S$-shaped conductivity we can define $\sigma$ as  the ratio between the bias and the dissipation
\begin{equation}
\label{sigma}
\sigma =  \frac{\gamma_R}{\alpha\left(  \gamma_R  \right)  }  \; .
\end{equation}
 					
\noindent The system of Equations (\ref{RSJ_S_general}) can be written:
\begin{eqnarray}
\label{RSJ_S_sigma}
 \frac{d\gamma_R}{d\tau} &=& \left\{
\frac{d \sigma \left(  \gamma_R  \right) }{d\gamma_R}
\right \}^{-1}
 \left[   \gamma_b -\gamma_R -  \sin(\phi)  \right] 					 \nonumber \\
 \frac{d\phi}{d\tau} &=& \frac{\gamma_R}{\alpha \left(  \gamma_R  \right)  }  \; .
\end{eqnarray}

\noindent The coefficient of the right hand side is the nonlinear differential conductance. Unfortunately for an $S$-shaped IV curve this is a term that vanishes in two points, see Fig. \ref{Lip}. For the model (\ref{sigma}), of the analytical expression of $d \sigma \left(  \gamma_R  \right) / d\gamma_R $  (see Eq.(\ref{sigma})), reads:
\begin{equation}
\label{sigmader}
\frac{d \sigma \left(  \gamma_R  \right) }{d\gamma_R}  =
   \frac{\alpha^2 \left(  \gamma_R  \right) }{\alpha\left(  \gamma_R  \right) - \gamma_R \frac{d\alpha\left(  \gamma_R  \right)}{d\gamma_R}}
  \; .
\end{equation}

\noindent The system becomes singular (non Lipschitzian) for two values of the current, where the differential conductance vanishes. Also, for negative values the system cannot be integrated
\begin{equation}
\label{sigma_non_lip}
\frac{d \sigma \left(  \gamma_R \right) }{d\gamma_R}  \le 0
 \; .
\end{equation}

\noindent Therefore, for the $S$-shaped case, one branch of the IV (and the corresponding power) cannot be reproduced, see Fig. \ref{FIG_Ibias_Sshape}.
A linear perturbation analysis of Eq.(\ref{RSJ_S_sigma}) around the singular points (\ref{sigma_non_lip}), can be accomplished setting $\gamma_R = \gamma' +\varepsilon (\tau)$:
\begin{eqnarray}
\label{cond_int}
\frac{d\phi}{d\tau} &=& \frac{\gamma '}{\alpha \left( \gamma ' \right)   } \equiv Z(\gamma ') 	\nonumber \\
\frac{dZ}{d \gamma'}(\gamma') \varepsilon \frac{d\varepsilon}{d\tau} &=&
\gamma_b - \gamma' -\sin\phi -\varepsilon
   \; .
\end{eqnarray}

Eqs. (\ref{cond_int}) predict that for small $\varepsilon$ (in the proximity of the singular point) the differential equations are not well defined. 
In the figures for the $S$-shaped resistor in fact squares denote the point implicitly defined by such condition, that well agree with the point where the differential equation cannot be numerically integrated.
Inserting an $S$-shaped resistor in $G\left(I_R\right)$ one gets in the simulations a typical result in Fig. \ref{FIG_Ibias_Sshape}.
Also for the $S$-shaped resistor we can assume a voltage bias, see 
Fig. \ref{FIG_Vbias_Sshape}. 
The normalized current is the same as in Eq. (\ref{Gbias}), that inserted into the system equation reads:
\begin{eqnarray}
\label{RSJ_S_voltage}
 \frac{d\gamma_R}{d\tau} &=&
\left[
\frac{\alpha^2 \left(  \gamma_R  \right) }{\alpha\left(  \gamma_R  \right) - \gamma_R \frac{d\alpha\left(  \gamma_R  \right)}{d\gamma_R}}
\right] \cdot \nonumber \\
& &   \left[
\frac{\alpha_0}{r_b}\left( \frac{V}{V_N}-\frac{d\phi}{d\tau}\right)  -\gamma_R -  \sin(\phi)
\right]  \nonumber \\
 \frac{d\phi}{d\tau} &=& \frac{\gamma_R}{\alpha\left(  \gamma_R  \right)}  \; .
\end{eqnarray}

\noindent The system reduces to the current bias when $\alpha_0/r_b\rightarrow 0$ and $V/V_N \rightarrow \infty$.

\section{Figure discussion}
\label{figure}
We note that there are no stable bias points between $A$ and $B$ in Figs. \ref{FIG_Ibias_Sshape} and \ref{FIG_Vbias_Sshape}. 
Increasing the bias current from zero in Fig. \ref{FIG_Ibias_Sshape} we first have a supercurrent until the current is $1$ where we have a switching to the lower branch of the IV curve. 
Increasing the bias current further we reach point $B$. 
Still increasing the bias current from $B$ we continue from point $A$. We note that the power ($I\cdot V$) dissipated in $A$ and $B$ are the same so the transition between $A$ and $B$ is rather unusual. 
There are no stable bias points between $A$ and $B$ and it is not a usual switching along a vertical or horizontal line. 

Figures \ref{FIG_Ibias_Sshape},\ref{FIG_Vbias_Sshape} include Josephson radiation in terms of voltage amplitude. 
In both Figs. \ref{FIG_Ibias_Sshape} and \ref{FIG_Vbias_Sshape} we also have unusual radiation patterns. 
Decreasing the bias current from above in Fig. \ref{FIG_Ibias_Sshape} we see the following: we observe radiation corresponding to the standard McCumber  curve, until we reach $A$. From here we continue down from $B$ and the radiation follows a new curve showing a peak deviating from the standard McCumber curve in region $I$. 
Increasing the bias current from zero we have the switching at $V=0$, but otherwise the radiation is the same as we had for decreasing bias current. 
Thus even if we have two branches of the radiation curves we do not have hysteresis in the usual sense. 
Figure \ref{FIG_Vbias_Sshape} with constant voltage bias shows essentially the same features.

\section{Summary}
$N$-type and $S$-type nonlinearities have been investigated in the McCumber curve for point Josephson junctions.

A  fictitious NDR, Gaussian shaped, perturbation of the standard McCumber dissipation is introduced to take into account the complicated electrodynamic interaction between stacks of JJ and the load. This allows us to obtian 
$N$-type and $S$-type nonlinearities that posses differential resistance. We found that the current-voltage characteristics of the full RSJ Josephson junction  model is very much influenced by a nonlinearly shaped McCumber curve leading to hysteresis and enhanced emitted radiation power in the region with negative differential resistance. Current-voltage curves were obtained for both current bias and voltage bias. 
More realistic high frequency schemes of the voltage bias and of the lumped elements could be attempted in further investigations. 
Our model governs a simple point Josephson junction, but the results can be applied to phase locked stacks of Josephson junctions, e.g. coupled via a common RLC cavity circuit.

\section*{Acknowledgements}
We thank COST program MP1201 Nanoscale Superconductivity: Novel Functionalities through Optimized Confinement of Condensate and Fields (NanoSC -COST) for partial support. GF work at the DTU was supported by the COST “Short Term Scientific Mission” COST-STSM-ECOST-STSM-MP1201-291013-036429. VP thanks INFN (Italy) for partial support.

\end{document}